\documentclass[aps,prc,twocolumn,groupedaddress]{revtex4-1}
\usepackage{graphicx}
\usepackage{dcolumn}
\usepackage{slashed}
\newcommand\be{\begin{equation}}
\newcommand\ee{\end{equation}}
\newcommand\ba{\begin{eqnarray}}
\newcommand\ea{\end{eqnarray}}

\begin{document}


\title{Nuclear Drell-Yan effect in a covariant model}

\author{C.\ L.\ Korpa}
\affiliation{Department of Theoretical Physics, University of P\'ecs, Ifj\'us\'ag \'utja 6, 7624 P\'ecs, Hungary}
\author{A.\ E.\ L.\ Dieperink}
\affiliation{Kernfysisch Versneller Instituut, Zernikelaan 25, NL-9747AA Groningen, The Netherlands}

\date{\today}

\begin{abstract}
We investigate effects of nuclear medium on antiquark distribution in nuclei applying the results
of a recently developed relativistically-covariant self-consistent model for the pion and the isobar.
We take into account Fermi motion including Pauli blocking and binding effects on the nucleons
and medium effects on the
isobar and pion leading to modest enhancement of the pion light-cone-momentum distribution in large
nuclei. As a consequence the Drell-Yan cross-section ratio with respect to the deuteron exceeds one
only for small values of the light-cone momentum.
\end{abstract}


\maketitle

\section{Introduction \label{Intro}}
Improving our understanding of the quark and gluon degrees of freedom
of nucleons bound in nuclei necessitates further efforts in spite of numerous investigations and successes \cite{Norton:2003}. 
The planned new nuclear Drell-Yan (DY) scattering experiment SeaQuest at Fermilab
(E-906) \cite{Reimer:2011zza,Reimer07,Reimer:2007zza} strongly motivates recent advancements in analysis of
nuclear parton distributions \cite{deFlorian:2011fp} as well as attempts directed at more reliable and
accurate calculation of experimentally observable quantities.
In line with the second objective is our aim to update and extend to
larger $x$ values previous calculations of nuclear DY effect presented in Refs.\
\cite{Diep97,Korpa99}. 
Furthermore, the physics interest is twofold: (i) to investigate the role of non-perturbative physics (pion cloud)
in the $\bar{u}-\bar{d}$ asymmetry in the nucleon, and (ii) to study possible anti-quark
enhancement due to presence of virtual mesons in the nuclear medium.
From symmetry properties of QCD we may infer that pion is one possible source of non-perturbative
quark-antiquark physics, in particular $u-d$ asymmetry.

In conventional models based upon meson exchange nuclear binding of nuclear matter comes
for about 50\% from (virtual) pions present in the nucleus. Can we observe these? Some indication for
pions is present in EMC effect enhancement around $x=0.1$, which can be ascribed to the fact that pions
(and heavier mesons) carry a fraction of the momentum sum rule. Can one see these pions more explicitly,
e.g.\ in the form of an enhancement of anti-quarks in the nucleus?
Anti-quarks can be probed directly in Drell-Yan scattering but
previous experiments \cite{Alde90} within the experimental uncertainty of about 10\% did not show a
nuclear enhancement; results of calculations varied strongly.

In practice one can distinguish two main types of theoretical interpretations of the classical
EMC effect: (i) in terms of nucleon as constituents which are bound but  not modified in the medium,
and (ii) in terms of off-shell nucleons with medium modified structure functions, e.g.\ through scalar
and vector fields acting on the quarks.
In the first category one has the non-relativistic models which use a computed spectral function that
accounts for large removal energy (50 MeV in nuclear matter) and Fermi motion due to correlations.
This approach can reproduce the observed slope of the reduction for $x<0.5$ in the
EMC ratio $2F_2^A(x)/AF_2^d(x)$ but not the behavior around the minimum in the ratio at $x=0.8$.
The latter seems to require ad-hoc off-shell effects \cite{Kulagin06}.
In the second category  \cite{Miller01, Mineo04}) one usually starts from the Walecka model
in the mean-field approximation which has a  small net binding effect (8 MeV per nucleon) and
hence yields a very small EMC effect, and then one adds the effect of external scalar and vector fields.
Since in the present study we are interested in the antiquark distribution for $x<0.4$ we rely on
the conventional convolution approach using a parameterized nucleon distribution, which has the
two above mentioned parameters related to the removal energy $\eta$ and the Fermi momentum.

In Sec.~II we study the antiquark distributions in the free nucleon and determine the off-shell
$\pi\text{NN}$ and $\pi\text{N}\Delta$ form-factors which lead to good description of the isovector
part of the proton antiquark distribution by the pion cloud. In Sec.~III we turn to general
discussion of the nuclear effects consisting of binding and Fermi motion of nucleons and modification
of the pion cloud. Detailed consideration of the medium effects on the nucleon's pion cloud
follows in Sec.~IV where expressions are derived for the pion light-cone distributions originating
from the $\pi\text{N}$ and $\pi\Delta$ states. We emphasize the careful treatment of the in-medium
delta baryon based on a complete relativistically covariant basis for its dressed propagator. Numerical
results for the in-medium pion distribution and DY cross-section ratios for nuclear targets relative
to the deuteron are presented and discussed in Sec.~V. Finally, Sec.~VI contains a summary of our results.
\section{Antiquarks in free nucleons}
Before turning to the nuclear case we want to investigate  whether the pion cloud approach we will
use in the medium can reproduce the observed flavor asymmetry in the free nucleon.
The distribution of antiquarks in the nucleon can be decomposed into a flavor-symmetric isoscalar part
(originating from
gluon splitting and possibly meson cloud) and a non-perturbative isovector meson-cloud contribution.The
latter can be considered to be the source of the $\bar{u}-\bar{d} $ asymmetry.
In addition to the pion cloud of the nucleon we include also the isobar with its pion cloud since it was
shown to give substantial contributions \cite{Kumano91}. In this way one can constrain the $\pi\text{NN}$ and
$\pi\text{N}\Delta$ form factors using the empirical antiquark u-d flavor asymmetry.
The physical free nucleon state is expressed approximately as
\be
|N\rangle = \sqrt{Z} |N\rangle_{\text{bare}}+ \alpha|N\pi\rangle + \beta |\Delta \pi\rangle.
\ee
Neglecting off-shell effects the light-cone momentum distribution of a quark with flavor $f$
in a proton
can be written as ($B=N,\Delta$):
\ba
q_f(x)&=& Z q_{f,{\text{bare}}}(x) +
\sum_{B,i} c_i\left[ \int_x^1 \frac{dy}{y} f^{B_i/N}(y)
q^{B_i}_{f,{\text{bare}}}(x/y)\right. \nonumber \\
 &+& \left. \int^1_x \frac{dy}{y} f^{\pi_i/N}(y) q^{\pi_i}(x/y)\right],
\label{pphys}
\ea
where $c_i$ ($i$ labels the charge states)
are the appropriate isospin Clebsch-Gordan coefficients, $q^{B_i}_{f,{\text{bare}}}(x)$
is the parton distribution in the bare $B_i$ baryon
and $q^\pi (x)$ is the pion parton distribution function.

Attributing the asymmetry in the $\bar u$ and $\bar d$ antiquark distributions to the nucleon
meson cloud we are concerned with the pion light-cone distribution in the nucleon which gets
contributions from final states with either nucleon or isobar:
\be
f^{\pi/N}(y)= f^{\pi N/N}(y)+ f^{\pi \Delta/N}(y).
\ee
The nucleon term was calculated by Sullivan \cite{Sullivan}:
\be
f^{\pi^0 N/N}(y)=\frac{g^2_{\pi NN}}{16\pi^2} y \int_{M^2y^2(1-y)}^\infty dt \frac{|F^{(\pi)}_{\pi NN}(t)|^2t}
{(t+m_\pi^2)^2},
\label{freepin}
\ee
with $y=(k_0+k_3)/M$ being the pion light-cone momentum
fraction, with $M$ the physical mass of the nucleon (as a convenient scale),
$F^{(\pi)}_{\pi NN} (t)$ is the off-shell form-factor of the $\pi NN$ vertex, while $g_{\pi NN}$ is the
$\pi^0 NN$ coupling. The free-pion propagator, $D_\pi^0$, appears
in the above expression in the form $(t+m^2_\pi)^{-1}$, where $t\equiv -q^2$ where $q$ denotes the pion
four-momentum.
The isobar contribution also plays an important role \cite{Kumano91} despite the kinematical
suppression coming from the isobar-nucleon mass difference. In Ref.~\cite{Kumano91} it was
calculated using the free isobar propagator, i.e.\ neglecting its width. A complete relativistically
covariant treatment of the isobar in vacuum and nuclear medium was introduced in Ref.~\cite{KD-2004}
and we use that formalism to take into account the vacuum width consistent with the measured
pion-nucleon scattering phase shift in the spin-3/2 isospin-3/2 channel. The full Lorentz structure of
the vacuum propagator of the Rarita-Schwinger field can be expressed in terms of 10 Lorentz scalar
functions \cite{Korpa-1997fk,KD-2004} which contains both spin-3/2 and spin-1/2 sectors \cite{KL-2006}.
However, using the convenient basis from Ref.~\cite{KD-2004} it
turns out that a single term, namely the (on mass shell) positive energy spin-3/2 contribution
gives the dominant contribution and all others (some terms in the propagator are identically zero)
are completely negligible.  In the notation of Ref.~\cite{KD-2004} this is the coefficient of the
projector sum $Q'^{\mu\nu}_{[11]}\equiv Q^{\mu\nu}_{[11]}+P^{\mu\nu}_{[55]}$ which we denote by
$G^{(Q')}_{[11]}(p)$. The pion light-cone distribution originating from the $\Delta\pi$ state then
can be expressed as:
\begin{eqnarray}
&&f^{\pi^-\Delta/N}(y)=\frac{yMg^2_{\pi N\Delta}}{6\pi^3}\,\int_{-\infty}^{-(My+m_\pi)}dp\,'_3\,
\int_0^{\infty}p\,'_\perp dp\,'_\perp \nonumber \\
&& \cdot F^{(\pi)}_{\pi N\Delta}(t)^2\, F^{(\Delta)}_{\pi N\Delta}(p\,')^2 
\frac{(M+p\cdot \hat p\,')\left(t+(k\cdot \hat p\,')^2\right)}{(t+m^2_\pi)^2}\nonumber \\
&&\cdot \text{Im}\,G^{(Q')}_{[11]}(p\,'),
\label{freepid}
\end{eqnarray}
where $p$ and $p\,'$ are the four-momenta of the nucleon and isobar, $t\equiv -(p-p\,')^2$ and
$F^{(\pi,\Delta)}_{\pi N\Delta}$ are the form factors of the $\pi N\Delta$ vertex.
The form factor
\be
F^{(\Delta)}_{\pi N\Delta}(p)=\exp\left[-\frac{p^2-(M+m_\pi)^2}{\Lambda^2}\right]
\ee
with $\Lambda=0.97\,$GeV (and $g_{\pi N\Delta}=20.2\,{\text{GeV}}^{-1}$) was used in Ref.~\cite{Korpa-1997fk}
and shown to give a good fit to the relevant pion-nucleon phase shift. For the $\pi NN$ and $\pi N \Delta$ off-shell form
factors which take into account the off-shell pion we take a dipole form:
\be
F^{(\pi)}_{\pi NX}(t)=\left( \frac{ \Lambda^2_{\pi X}-m^2_\pi}{ \Lambda^2_{\pi X}+t} \right)^2,
\ee
with $X$ standing for $N$ or $\Delta$.

\begin{figure}
\includegraphics[width=9cm]{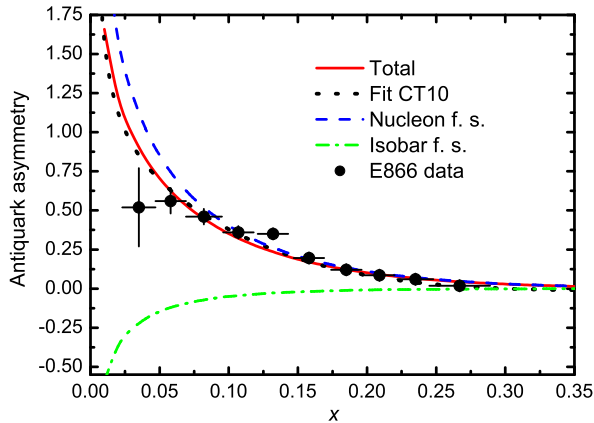}
\caption{(Color online) The pion-cloud result for the $\bar d-\bar u$ asymmetry in free proton (solid line) compared to
the difference of the proton $\bar d$ and $\bar u$ distributions (dot line) from the fit CT10 \cite{Lai-2010} and
data points by Fermilab E866/NuSea Collaboration \cite{E866-98}.
The isobar contribution
(dash-dot line) is negative and much smaller than the nucleon term (dash line). Used parameter values:
$\Lambda^{(\pi)}_{\pi N}=0.95\,$GeV and $\Lambda^{(\pi)}_{\pi\Delta}=0.75\,$GeV.
\label{duasymf}}
\end{figure}

In order to calculate the $\bar d-\bar u$ distribution for the free proton we assume that the pion sea is
isospin symmetric leaving only the contribution of valence distributions. The final state with nucleon
contributes through the presence of $\pi^+$ with distribution $2f^{\pi^0 N/N}$, while the isobar final state
can have also a $\pi^+$ with isospin weight $1/3$ or a $\pi^-$ with minus sign (because of the pion valence $\bar u$
distribution) with respect to $f^{\pi^-\Delta/N}$ giving in total:
\be
(\bar d-\bar u)_p(x)=\int_x^1 \frac{dy}{y} \left( 2f^{\pi^0 N/N}(y)-\frac{2}{3}f^{\pi^-\Delta/N}(y)\right)
q_v^\pi(x/y),
\label{duasym}
\ee
with $q_v^\pi(x)$ denoting the valence parton distribution of charged pion.
\begin{figure}
\includegraphics[width=9cm]{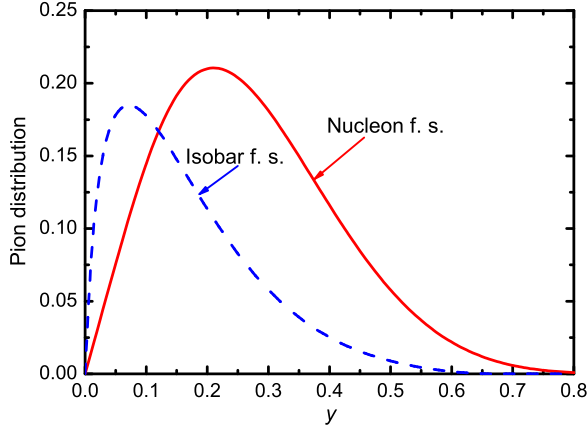}
\caption{(Color online) Pion distribution in the free proton: $f^{\pi^0 N/N}(y)$ shown by solid line and $f^{\pi^-\Delta/N}(y)$
by dash line.
\label{freepidis}}
\end{figure}
In Fig.~\ref{freepidis} we show the pion distributions with nucleon final state (solid line) and isobar final
state (dash line). For the form-factor cutoff we used the following values: $\Lambda^{(\pi)}_{\pi N}=0.95\,$GeV and
$\Lambda^{(\pi)}_{\pi\Delta}=0.75\,$GeV. The bare-nucleon probability then takes the value $Z=0.69$ which suggests
that higher-order terms with more than one pion do not contribute significantly. 
The calculated value for the $\bar d-\bar u$ asymmetry in the free proton
is shown in Fig.~\ref{duasymf} by solid line and compared to the result using the $\bar u$ and $\bar d$ fits
CT10 \cite{Lai-2010} (dot line). Also shown are separately the contributions from the nucleon final state (dash line) and
isobar final state (dash-dot line).
These results are quite similar to the $\bar d-\bar u$ obtained in Ref.~\cite{Meln98}, although there the
infinite-momentum-frame formalism was used with suitably adjusted values of the $\pi\text{N}$ and $\pi\Delta$
form factors.
\section{Nuclear effects}
First calculations of the nuclear Drell-Yan process \cite{Bick84} suggested an enhancement coming
from the medium modification of the pion cloud. However, the experimental data \cite{Alde90} did not
show that enhancement within a 10\% uncertainty. Later on other groups reported more detailed
calculations of the Drell-Yan ratio with a large variation in results as shown in
Refs.~\cite{Reimer:2011zza,Reimer07,Reimer:2007zza}.
Here we consider the ratio of the cross sections of proton-nucleus and proton-deuteron scattering,
\begin{equation}
R_{A/d}= \frac{2}{A}\frac {d\sigma^{pA}/dx_1dx_2} {d\sigma^{pd}/dx_1dx_2}.
\end{equation}
where $A$ denotes the nucleus and its nucleon number. We specialize for the case of isoscalar targets for
which the cross-section ratio becomes:
\begin{widetext}
\begin{equation}
R_{A/d}=
\frac
{\sum_f e^2_f
\left\{ q^p_f(x_1)\left[ \overline{q_f^{p/A}}(x_2)+\overline{q_f^{n/A}}
(x_2)
\right] +\overline{q_f^p}(x_1)\left[ q^{p/A}_f(x_2)+q^{n/A}_f(x_2)\right]
\right\} }
{\sum_f e^2_f
\left\{ q^p_f(x_1)\left[ \overline{q_f^p}(x_2)+\overline{q_f^n}(x_2)
\right] +\overline{q_f^p}(x_1)\left[ q^p_f(x_2)+q^n_f(x_2)\right]
\right\} }. \label{ratio}
\end{equation}
\end{widetext}
In the case when $x_1$ is large, say $x_1>0.3$ the second term in the numerator becomes negligible
and only medium effect on the antiquarks plays a role.

\begin{figure}
\includegraphics[width=9cm]{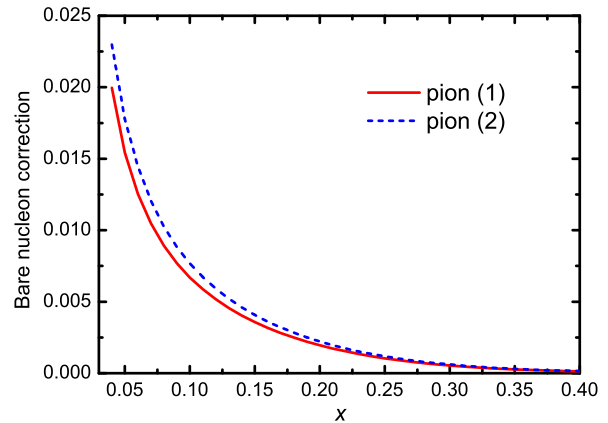}
\caption{(Color online) Sum of the second, third, fourth and fifth terms in Eq.~(\ref{pnmed1}) devided by the antiquark 
distribution of the free ``isoscalar" nucleon. Solid line is for the pion parameter set (1) and dash line for parameter set (2).
\label{termadd}}
\end{figure}

The (anti-)quark distribution in the medium can be modified in two ways, (i) through Fermi motion
and binding of the nucleon, and (ii) modification of the nucleon's pion cloud.
To establish the connection to the (anti-)quark distribution of the free nucleon we use Eq.~(\ref{pphys})
which for the free proton gives (with isobar terms not written out for brevity):
\begin{eqnarray}
&&q_f^p(x)=Z q^p_{f,{\text{bare}}}(x)+
\frac{1}{3}\int_x^1 \frac{dy}{y} f^{N/N}(y)
\left[ q^p_{f,{\text{bare}}}(x/y)+\right. \nonumber \\
&&\left. 2q^n_{f,{\text{bare}}}(x/y)\right]
+\int_x^1 \frac{dy}{y} f^{\pi^0N/N}(y)
\left[ q_f^{\pi^0}(x/y)+2q_f^{\pi^+}(x/y)\right],\label{qfree}
\end{eqnarray}
with
\begin{equation}
Z \equiv 1- \int_0^1 dy f^{N/N}(y)=1-3\int_0^1 dy f^{\pi^0N/N}(y),
\label{Z}
\end{equation}
where the last equality expresses flavor-charge conservation.
Similarly, the quark distribution for the nuclear proton can be written:
\begin{eqnarray}
&&\tilde{q}_f^p(x)=Z_A q^p_{f,{\text{bare}}}(x)+
\frac{1}{3}\int_x^A \frac{dy}{y} f^{N/A}(y)
\left[ q^p_{f,{\text{bare}}}(x/y)+\right. \nonumber \\
&&\left. 2q^n_{f,{\text{bare}}}(x/y)\right]
+\int_x^A \frac{dy}{y} f^{\pi^0N/A}(y)
\left[ q_f^{\pi^0}(x/y)+2q_f^{\pi^+}(x/y)\right],\label{qmed}
\end{eqnarray}
where isospin-symmetric nuclear medium was assumed.
Adding the difference of the left-hand side and right-hand side of (\ref{qfree}) to
the right-hand side of (\ref{qmed}) and repeating the same
procedure for the neutron we obtain
\begin{eqnarray}
&&\tilde{q}^p_f(x)+\tilde{q}^n_f(x)=
q^p_f(x)
+q^p_{f,{\text{bare}}}(x)\int_0^1 f^{N/N}(y)dy-\nonumber \\ 
&&\int_x^1 \frac{dy}{y}
f^{N/N}(y)q^p_{f,{\text{bare}}} (x/y)
-q^p_{f,{\text{bare}}}(x)\int_0^A f^{N/A}(y)dy+\nonumber \\
&&\int_x^A \frac{dy}{y}
f^{N/A}(y)q^p_{f,{\text{bare}}} (x/y)
+(p\rightarrow n)\nonumber \\
&&+2\int_x^A \frac{dy}{y} \left[f^{\pi^0N/A}(y)-
f^{\pi^0N/N}(y)\right] \nonumber \\
&& \cdot\left[ q^f_{\pi^0}(x/y)+
q^f_{\pi^+}(x/y)+
q^f_{\pi^-}(x/y)
\right],\label{pnmed1}
\end{eqnarray}
where we used that
\[
Z_A \equiv 1- \int_0^1 dy f^{N/A}(y).
\]
To proceed with the above expression one needs the bare antiquark distributions which could be determined from
Eq.~(\ref{qfree}) and its analog for the neutron. A much simpler, though approximate procedure is just
to subtract the meson-cloud contribution form the antiquark distribution of the physical nucleon. Indeed, using
the fact that antiquark distributions at small $x$ behave as $1/x$ one can confirm that the first and second term
on the right-hand side of (\ref{qfree}) combine to give the bare distribution if one adds to (\ref{qfree}) its
neutron analog. Using the same argument about the small $x$ behavior of antiquark distributions one can
establish an approximate cancellation of the second and third as well as the fourth and fifth terms on the right-hand side
of Eq.~(\ref{pnmed1}) and of the corresponding terms involving the neutron.
This simplification was used in our previous work \cite{Diep97}, but in the present calculation
we want to take into account these contributions with the bare antiquark distributions determined by the above mentioned
subtraction of the pion contribution from the physical distribution. In Fig.~\ref{termadd} we show the proton-neutron average of the
sum of the second,
third, fourth and fifth terms in Eq.~(\ref{pnmed1}) divided by the antiquark
distribution of the free ``isoscalar" nucleon for two used pion parameter sets.
For the distribution of the bare nucleon in the free nucleon we use the relation $f^{N/N}(y)=3f^{\pi^0N/N}(1-y)$ which follows
from the probabilistic interpretation of these functions \cite{Meln98} and its analog for the in-medium case,
$f^{N/A}(y)=3f^{\pi^0N/A}(1-y)$. We remark that the latter relationship can only be approximate since the support of the 
pion in-medium distribution is not strictly limited by value one, but in view of the similarity of the pion distributions
in the two cases it should be a reasonable approximation for the estimate of a small effect.
We observe that this contribution is indeed quite
small as expected from the form of the nucleon antiquark distribution and pion (as well as related nucleon) light-cone-momentum
distributions. The parameter set (1) is: $M_*=0.89\,\text{GeV}, \Sigma^v_N=0, \Sigma^s_\Delta=-0.1\,\text{GeV},
\Sigma^v_\Delta=0, g'_{11}=0.9, g'_{12}=0.3, g'_{22}=0.3$, while the set (2) is given by: $M_*=0.89\,\text{GeV}, \Sigma^v_N=0,
\Sigma^s_\Delta=-0.05\,\text{GeV},\Sigma^v_\Delta=0, g'_{11}=1.0, g'_{12}=0.4, g'_{22}=0.3$; where $M_*$ is the mean-field shifted nucleon
mass, $\Sigma^v_N$ the energy shift of the nucleon, $\Sigma^s_\Delta$ and $\Sigma^v_\Delta$ the delta's mean-field shifts and $g'_{ij}$
the Migdal four-fermion interaction parameters.

Introducing a shorter notation for the bare nucleon contribution to the in-medium antiquark distribution:
\begin{widetext}
\begin{eqnarray}
q^{p+n}_{f,{\text{bare}}}(x)&=&q^p_f(x)
+q^p_{f,{\text{bare}}}(x)\int_0^1 f^{N/N}(y)dy-\int_x^1 \frac{dy}{y}
f^{N/N}(y)q^p_{f,{\text{bare}}} (x/y)\nonumber \\
&-& q^p_{f,{\text{bare}}}(x)\int_0^A f^{N/A}(y)dy+\int_x^A \frac{dy}{y}
f^{N/A}(y)q^p_{f,{\text{bare}}} (x/y)+(p\rightarrow n),
\end{eqnarray}
we can finally write the sum of in-medium proton and neutron antiquark distribution as
\be
q^{p/A}_f(x)+q^{n/A}_f(x)=\int_x^A \frac{dy}{y} f^N_{Fb}(y)
q^{p+n}_{f,{\text{bare}}}(x/y)
+
2\int_x^A \frac{dy}{y} \left[f^{\pi^0/A}(y)-
f^{\pi^0/N}(y)\right] \left[ q^f_{\pi^0}(x/y)+
q^f_{\pi^+}(x/y)+
q^f_{\pi^-}(x/y)
\right].       \label{qmedpn}
\ee
\end{widetext}
The convolution with $f^N_{Fb}(z)$ takes into account Fermi motion and binding effects on the in-medium
nucleons. For the function $f^N_{Fb}(z)$ we take the result of Birse \cite{Birse93}:
\begin{equation}
f^N_{Fb}(z)=\frac{3}{4\epsilon^3}\left[ \epsilon^2-
(z-\eta)^2 \right] \Theta \!\left( \epsilon-|z-\eta| \right),
   \label{fermirel}
\end{equation}
where $\epsilon\equiv p_F/M$, $\eta$ is a parameter with value slightly below one which takes
into account the nuclear binding and $\Theta(x)$ is the unit step function.
\begin{figure}
\includegraphics[width=9cm]{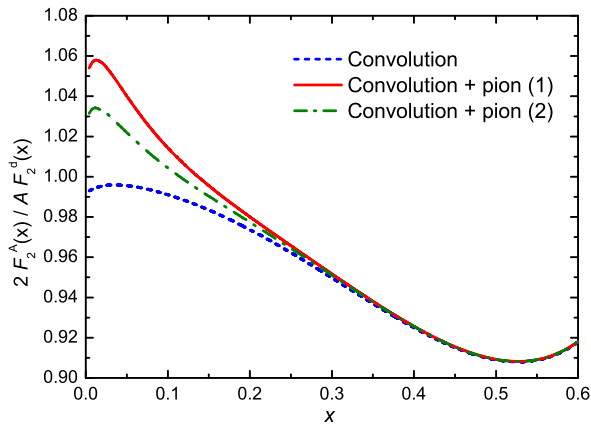}
\caption{ (Color online) The ratio of $F_2(x)$ per nucleon for isospin symmetric nuclear medium with parameters:
Fermi momentum $p_F=250\,$MeV, $\eta=0.97$, and deuteron. Pion contribution with parameter sets (1) and (2) is included
in the results shown by solid and dash-dot lines.}
\label{f2ratio}
\end{figure}
In Fig.~\ref{f2ratio} we show the ratio of the $F_2(x)$ structure functions for the isospin symmetric nuclear matter and the
deuteron. We assume negligible medium effects in the deuteron and for the nuclear medium use the convolution model
with light-cone distribution (\ref{fermirel}) and parameters $p_F=250\,$MeV and $\eta=0.97$. In this way one can reproduce
the negative slope in the classical EMC effect for $0.1<x<0.5$ as shown for example in Ref.~\cite{Arnold-84}.
The experimental enhancement observed around $x=0.1$ can
be attributed to the pion enhancement as shown in the figure for the two parameter sets (1) and (2) used also for the
plots in Fig.~\ref{termadd} and given above. The pion enhancement term was calculated by the convolution of the in-medium
pion light-cone distribution enhancement relative to the free nucleon and the pion $F_2$ distribution.
Note that expected shadowing effects would lead to decrease of the nuclear cross section for $x\leq 0.05$.

\section{Antiquarks in bound nucleons: pion contribution}
We now turn to consideration of the pion contribution to antiquark distributions in nucleons bound in large nuclei which
we model by an infinite system with appropriate average nuclear density.

For corresponding pion properties in the nuclear matter we use the results of a fully covariant
self-consistent model developed in Ref.~\cite{KLR-2009}. Compared to the case of the free nucleon
the nuclear environment changes the pion propagator appearing in the Sullivan formula (\ref{freepin})
and renormalizes the $\pi\mbox{NN}$ as well as the $\pi\mbox{N}\Delta$ vertices through nucleon-nucleon
correlations modeled by the Migdal four-fermion interactions \cite{Migdal-1978}. Nucleon properties
are also affected and we take into account the binding effects through mean-field mass and energy
shifts consistent with approach in Ref.~\cite{KLR-2009}.

\begin{figure}
\includegraphics[width=7cm]{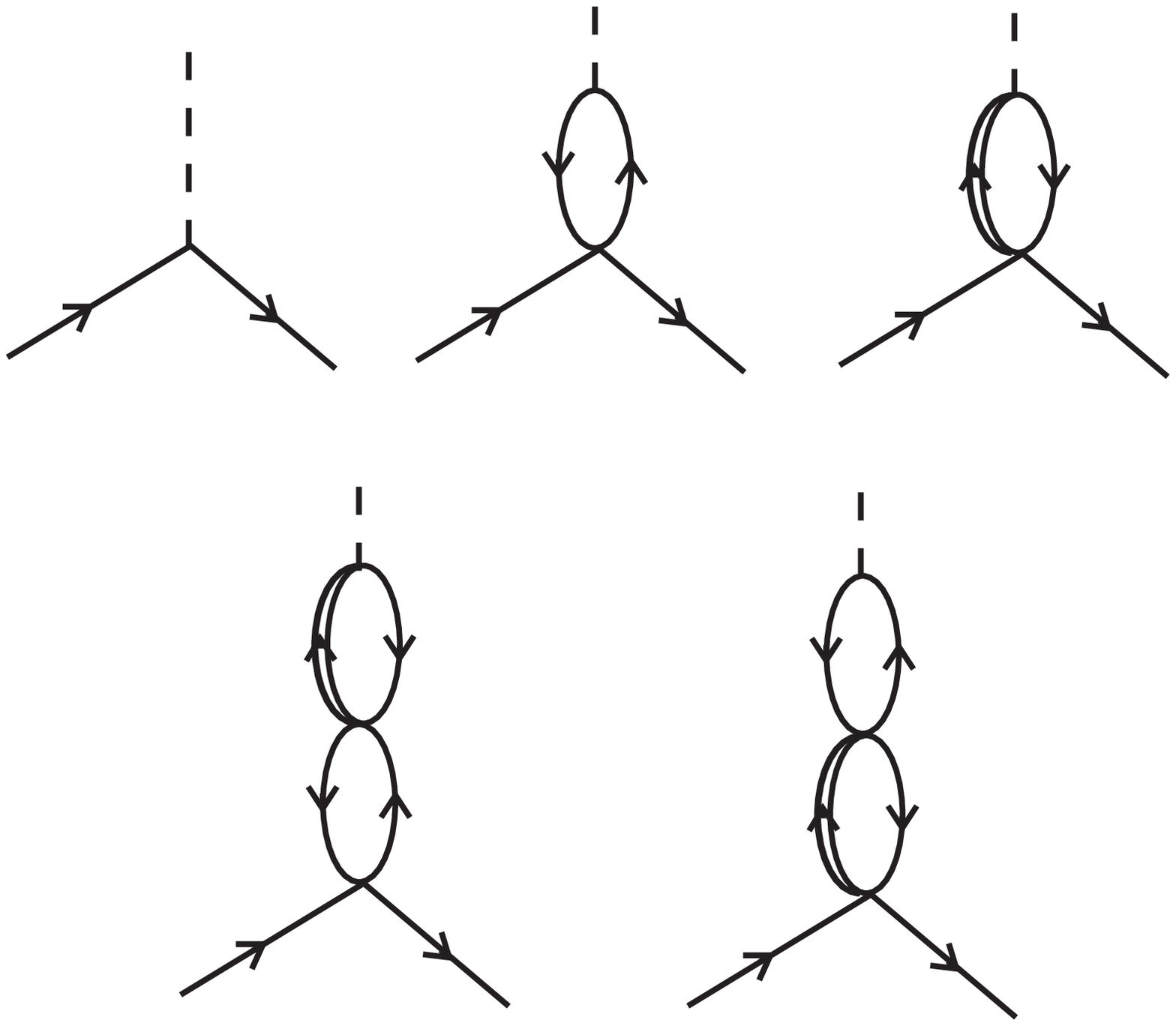}
\caption{Types of diagrams contributing to the in-medium pion distribution with outgoing nucleon. The same types
appear also with outgoing delta baryon. The dash line denotes the dressed
pion propagator, while the solid and the double line correspond to nucleon and delta.
\label{diagrams}}
\end{figure}

The inclusion of the dressed pion propagator is straightforward but the dressing of the $\pi\mbox{NN}$
and the $\pi\mbox{N}\Delta$ vertices requires summation of nucleon-hole and delta-hole bubbles. The types
of relevant diagrams are shown in Fig.~\ref{diagrams}.
For resummation of these diagrams
we use the relativistically covariant formalism introduced in Ref.~\cite{Lutz-2003} and applied
for pion and delta self energy calculation in Ref.~\cite{KLR-2009}. In the present case it concerns a different type of
contribution which for the nucleon in the final state can be written:
\begin{equation}
K_N=\bar u(p') \gamma_5 \gamma^\mu u(p)\cdot \Pi_{\mu\nu}(q) q^\nu,
\label{ndiag}
\end{equation}
where $q=p-p'$ is the pion 4-momentum, $u(p)$ is the nucleon in-medium spinor (with mean-field shifts
of mass and energy) and $\Pi_{\mu\nu}(q)$ is the resummed contribution of nucleon-hole and delta-hole
loops. Using the decomposition of nucleon-hole and delta-hole loops \cite{KLR-2009}:
\begin{eqnarray}
&& \Pi_{\mu \nu}^{(N h)}(q) = \sum_{i,j=1}^2\,
\Pi_{ij}^{(N h)}(q)\,L_{\mu \nu}^{(ij)}(q)+ \Pi_{\,T}^{(N h)}(q)\,T_{\mu \nu}(q) \,,
\nonumber\\
&& \Pi_{\mu \nu}^{(\Delta h)}(q) = \sum_{i,j=1}^2\,
\Pi_{ij}^{(\Delta h)}(q)\,L_{\mu \nu}^{(ij)}(q)
+ \Pi_{\,T}^{(\Delta h)}(q)\,T_{\mu \nu}(q)\,,
\label{def-decom-pi}
\end{eqnarray}
where $L_{\mu \nu}^{(ij)}(q)$ and $T_{\mu \nu}(q)$ have projector properties, the nonzero contribution
involving one particle-hole (nucleon-hole or delta-hole) loop comes from
\begin{eqnarray}
\Pi^{(1)}_{\mu\nu}&=&g'_{11}\left( \Pi_{11}^{(N h)}(q)\,L_{\mu \nu}^{(11)}(q)+
\Pi_{21}^{(N h)}(q)\,L_{\mu \nu}^{(21)}(q)\right) \nonumber \\
&+&
g'_{12}\left( \Pi_{11}^{(\Delta h)}(q)\,L_{\mu \nu}^{(11)}(q)+
\Pi_{21}^{(\Delta h)}(q)\,L_{\mu \nu}^{(21)}(q)\right),
\label{1st}
\end{eqnarray}
where $g'_{11}, g'_{12},g'_{22}$ are the usual Migdal parameters \cite{KLR-2009} where index 1 refers to the nucleon and 
index 2 to the delta and we took into account that:
\begin{equation}
 L_{\mu \nu}^{(12)}(q)\cdot q^\nu=L_{\mu \nu}^{(22)}(q)\cdot q^\nu=T_{\mu \nu}(q)\cdot q^\nu=0.
\end{equation}
In order to perform the summation involving arbitrary number of nucleon-hole or delta-hole loops it
is convenient to introduce the following matrices \cite{Lutz-2003}:
\begin{widetext}
\begin{eqnarray}
&& g^{(L)} = \left(
\begin{array}{llll}
g_{11}' & 0 & g_{12}' & 0 \\
0 & g_{11}' & 0 & g_{12}'  \\
g_{12}' & 0 & g_{22}' & 0 \\
0 & g_{12}' & 0 & g_{22}'
\end{array}
\right) \,, \quad \!\Pi^{(L)}= \left(
\begin{array}{llll}
\Pi_{11}^{(N h)} & \Pi_{12}^{(N h)} & 0 & 0 \\
\Pi_{21}^{(N h)} & \Pi_{22}^{(N h)} & 0 & 0  \\
0 & 0 & \Pi_{11}^{(\Delta h)} & \Pi_{12}^{(\Delta h)} \\
0 & 0 & \Pi_{21}^{(\Delta h)} & \Pi_{22}^{(\Delta h)}
\end{array}
\right)\,. \label{def-matrix}
\end{eqnarray}
The lowest order contribution (\ref{1st}) can then be written as:
\begin{eqnarray}
\Pi^{(1)}_{\mu\nu}&=&[(\Pi^{(L)}g^{(L)})_{11}+(\Pi^{(L)}g^{(L)})_{31}]\,L_{\mu \nu}^{(11)}(q)
 \nonumber \\
&+&
[(\Pi^{(L)}g^{(L)})_{21}+(\Pi^{(L)}g^{(L)})_{41}]\,L_{\mu \nu}^{(21)}(q).
\label{1st1}
\end{eqnarray}
Higher order terms are accounted for by taking appropriate matrix elements of products of
$(\Pi^{(L)}g^{(L)})$ matrices and the summation of terms with arbitrary number of loops is
simply achieved by replacing $(\Pi^{(L)}g^{(L)})$ in (\ref{1st1}) by
$\Pi^{(L)}g^{(L)}\,(1-\Pi^{(L)}g^{(L)})^{-1}$ leading to:
\begin{eqnarray}
\Pi_{\mu\nu}&=&\left( [\Pi^{(L)}g^{(L)}\,(1-\Pi^{(L)}g^{(L)})^{-1}]_{11}+
[\Pi^{(L)}g^{(L)}\,(1-\Pi^{(L)}g^{(L)})^{-1}]_{31}\right) \,L_{\mu \nu}^{(11)}(q)
 \nonumber \\
&+&
\left( [\Pi^{(L)}g^{(L)}\,(1-\Pi^{(L)}g^{(L)})^{-1}]_{21}+
[\Pi^{(L)}g^{(L)}\,(1-\Pi^{(L)}g^{(L)})^{-1}]_{41}\right) \,L_{\mu \nu}^{(21)}(q).
\label{pi-summed}
\end{eqnarray}
Adding $g_{\mu\nu}$ to $\Pi_{\mu\nu}$ in (\ref{ndiag}) we obtain the full contribution of the
diagram with nucleon final state. Squaring its absolute value and performing summation over the
spin projections of the nucleon in the final state and averaging for the nucleon in the initial
state we obtain:
\begin{eqnarray}
\frac{1}{2}\,\sum_{s,s'} |K_N|^2&=&
A_{qq}\cdot \left(2 (p-\Sigma_N^v u)\cdot q\,(p'-\Sigma_N^v u)\cdot q -
q^2\, [M_*^2+(p'-\Sigma_N^v u)\cdot (p-\Sigma_N^v u)]\right) \nonumber \\
&+&
2A_{qu}\cdot \left((p-\Sigma_N^v u)\cdot u \,(p'-\Sigma_N^v u)\cdot q +(p'-\Sigma_N^v u)\cdot u \,
(p-\Sigma_N^v u)\cdot q\right. \nonumber \\
&-& \left.  q\cdot u \,[M_*^2+(p'-\Sigma_N^v u)\cdot (p-\Sigma_N^v u)] \right) \nonumber\\
&+& A_{uu}\cdot\left(2(p-\Sigma_N^v u)\cdot u\,(p'-\Sigma_N^v u)\cdot u -M_*^2-(p'-\Sigma_N^v u)\cdot (p-\Sigma_N^v u)\right),
\label{nsum}
\end{eqnarray}
where $u$ is the 4-velocity of the medium (implicitly present also in Eqs.~(\ref{ndiag})$-$(\ref{pi-summed})),
$M_*=M_N+\Sigma_N^s$, $\Sigma_N^s$ and $\Sigma_N^v$ are the nucleon mean-field mass and energy shifts. The
factors $A_{qq}, A_{qu}, A_{uu}$ are given by:
\begin{eqnarray}
A_{qq}&=&2\left|1+A+q\cdot u\,B/\sqrt{q^2-(q\cdot u)^2}\right|^2,\nonumber \\
A_{qu}&=&-2\, {\text{Re}}\left[ q^2 B/\sqrt{q^2-(q\cdot u)^2}(1+\bar A -q\cdot u\,\bar{B}/\sqrt{q^2-(q\cdot u)^2})\right]
\nonumber \\
A_{uu}&=& 2\left|q^2\,B/\sqrt{q^2-(q\cdot u)^2}\right|^2,
\end{eqnarray}
where the bar denotes complex conjugation,
\begin{eqnarray}
A &\equiv &[\Pi^{(L)}g^{(L)}\,(1-\Pi^{(L)}g^{(L)})^{-1}]_{11}+
[\Pi^{(L)}g^{(L)}\,(1-\Pi^{(L)}g^{(L)})^{-1}]_{31}, \nonumber \\
B &\equiv &[\Pi^{(L)}g^{(L)}\,(1-\Pi^{(L)}g^{(L)})^{-1}]_{21}+
[\Pi^{(L)}g^{(L)}\,(1-\Pi^{(L)}g^{(L)})^{-1}]_{41},
\end{eqnarray}
\end{widetext}
and we used that:
\begin{eqnarray}
L_{\mu \nu}^{(11)}(q)\,q^\nu&=&q_\mu,\nonumber \\
L_{\mu \nu}^{(21)}(q)\,q^\nu&=&\frac{q\cdot u}{\sqrt{q^2-(q\cdot u)^2}}\,q_\mu-\frac{q^2}{\sqrt{q^2-(q\cdot u)^2}}\,u_\mu.
\end{eqnarray}

To compute the pion light-cone distribution per nucleon in the medium we integrate over incoming nucleons
in the Fermi sea and outgoing ones above the Fermi sea, restricting the pion light-cone momentum fraction to
the specified value by inserting a delta function and finally divide by the nucleon density. The final expression
obtained is:
\begin{eqnarray}
&& f_N(y)=3My\left(\frac{f_N}{m_\pi}\right)^2 \frac{1}{32\pi^3 p_F^3}\int_{-p_F}^{p_F}dp_3
\int_{0}^{\sqrt{p_f^2-p_3^2}}p_\perp dp_\perp \nonumber \\
&& \cdot \int_{p'^{\text{min}}_\perp} ^\infty p'_\perp dp'_\perp 
\int_0^{2\pi} d\vartheta \,\frac{1}{2b}\,\sum_{s,s'} |K_N|^2\nonumber \\ 
&&\cdot \left|F^{(\pi)}_{\pi NN}(-q^2)\,
D_\pi(q) \right|^2,
\label{pidismedn}
\end{eqnarray}
where $D_\pi(q)$ is the in-medium dressed pion propagator, $b\equiv My-p_3-\sqrt{M_*^2+p_3^2+p_\perp^2}$,
$p'^{\text{min}}_\perp=\sqrt{2b\sqrt{M_*^2+p_F^2}-M_*^2-b^2}$, $\vartheta$ is the angle between $\vec p_\perp$ and
$\vec p\,'_\perp$, and the $\pi NN$ form factor $F^{(\pi)}_{\pi NN}(-q^2)$ was also included.

The contribution coming from the delta baryon in the final state is made more involved by the complicated
structure of the in-medium delta propagator \cite{KD-2004}. However, considerable simplification can be
achieved by including only the two dominant contributions in the convenient relativistically covariant
decomposition since the imaginary part of the other components is typically two orders of magnitude smaller at
nuclear saturation density \cite{Korpa-2012}. The dominant contributions come from the $Q^{\mu\nu}_{[11]}$ and
$P^{\mu\nu}_{[55]}$ terms which were degenerate in the free delta case, but are different in the medium \cite{KD-2004}.
Summation of particle-hole loops dressing the $\pi N\Delta$ vertex is analogous to the $\pi NN$ case, with only
difference being in the relevant matrix elements of $\Pi^{(L)}g^{(L)}\,(1-\Pi^{(L)}g^{(L)})^{-1}$,
replacing the expression (\ref{pi-summed}) with:
\begin{widetext}
\begin{eqnarray}
\Pi^\Delta_{\mu\nu}&=&\left( [\Pi^{(L)}g^{(L)}\,(1-\Pi^{(L)}g^{(L)})^{-1}]_{13}+
[\Pi^{(L)}g^{(L)}\,(1-\Pi^{(L)}g^{(L)})^{-1}]_{33}\right) \,L_{\mu \nu}^{(11)}(q)
 \nonumber \\
&+&
\left( [\Pi^{(L)}g^{(L)}\,(1-\Pi^{(L)}g^{(L)})^{-1}]_{23}+
[\Pi^{(L)}g^{(L)}\,(1-\Pi^{(L)}g^{(L)})^{-1}]_{43}\right) \,L_{\mu \nu}^{(21)}(q).
\label{pid-summed}
\end{eqnarray}
The expression analogous to (\ref{nsum}) in this case takes the form:
\begin{equation}
\frac{1}{2}\,\sum_{s,s'} |K_\Delta|^2=\frac{1}{2}\,{\text{Tr}}\left[
(\slashed p -\Sigma_N^v \slashed u+M_*){\text{Im}}\,G^{\mu\nu}(p')\right]\left(
g_{\mu\alpha}g_{\nu\beta}+\Pi^\Delta_{\mu\alpha}\bar \Pi^\Delta_{\nu\beta}\right) q^\alpha q^\beta,
\label{dsum}
\end{equation}
where ${\text{Im}}\,G_{\mu\nu}(p')$ denotes the imaginary part of the in-medium $\Delta$ propagator for
which we take the dominant contribution given in the basis used in Ref.~\cite{KD-2004} by just two terms:
\begin{equation}
G^{\mu\nu}(p')=Q^{\mu\nu}_{[11]}(p')\,G^{(Q)}_{[11]}(p')+P^{\mu\nu}_{[55]}(p')\,G^{(P)}_{[55]}(p').
\end{equation}
In this way the expression (\ref{dsum}) takes the form:
\begin{eqnarray}
\frac{1}{2}\,\sum_{s,s'} |K_\Delta|^2 &=&
\left[ A_{qq}^{(\Delta)}\,c_{qq}^{(Q)} +2 A_{qu}^{(\Delta)}\,c_{qu}^{(Q)}+A_{uu}^{(\Delta)}\,c_{uu}^{(Q)} \right]\,
{\text{Im}}\,G^{(Q)}_{[11]}(p') \nonumber \\
&+&
\left[ A_{qq}^{(\Delta)}\,c_{qq}^{(P)} +2 A_{qu}^{(\Delta)}\,c_{qu}^{(P)}+A_{uu}^{(\Delta)}\,c_{uu}^{(P)} \right]\,
{\text{Im}}\,G^{(P)}_{[55]}(p'),
\label{dsum1}
\end{eqnarray}
where the expressions for $A_{qq}^{(\Delta)}, A_{qu}^{(\Delta)},A_{uu}^{(\Delta)},c_{qq}^{(Q)},c_{qu}^{(Q)},
c_{uu}^{(Q)},c_{qq}^{(P)},c_{qu}^{(P)},c_{uu}^{(P)}$ are given in the Appendix.
The pion light-cone momentum distribution stemming from the process with the nucleon emitting a pion and a
delta baryon is analogous to expression (\ref{pidismedn}) and reads:
\begin{eqnarray}
f_\Delta(y)&=&3My\left(\frac{f_\Delta}{m_\pi}\right)^2 \frac{1}{32\pi^4 p_F^3}\int_{-p_F}^{p_F}dp_3
\int_{0}^{\sqrt{p_f^2-p_3^2}}p_\perp dp_\perp
\int_{-\infty} ^\infty dp'_3 \nonumber \\
&&\cdot \int_{0}^{\infty}p'_\perp dp'_\perp \int_0^{2\pi} d\vartheta \,\frac{1}{2}\,\sum_{s,s'}
|K_\Delta|^2\,\left|F^{\Delta}_{\pi N\Delta}(p')\,F^{(\pi)}_{\pi N\Delta}(-q^2)\,
D_\pi(q)\right|^2.
\label{pidismedd}
\end{eqnarray}
\end{widetext}
We checked by explicit numerical calculation that both Eqs.~(\ref{pidismedn}) and (\ref{pidismedd}) have
the correct low-density limit, i.e.\ reproduce the free nucleon and delta results.
\section{Numerical results and discussion}
For the computation of in-medium pion and isobar properties we rely on the recently developed
relativistically covariant self-consistent model presented in Ref.~\cite{KLR-2009} and used for
the nuclear photoabsorption calculation in the isobar region in Ref.~\cite{RLK-2009}. For the medium
computation we use the same $\pi\text{NN}$ and $\pi\text{N}\Delta$ form factors as in the vacuum one and
include them in the model of Ref.~\cite{KLR-2009}. The values of the Migdal $g'$ parameters which model
the short-range nucleon and isobar correlations were taken in the range preferred by the results of Ref.~\cite{RLK-2009};  in this  work
 a good description of the nuclear photo-absorption cross section in the isobar region was obtained. Binding effects
for the nucleon are taken into account by the effective (mean-field)  mass $M_*$ and the energy shift $\Sigma_N^v$.
A consequence of the use of the  mean-field approximation is a reduction of the in-medium pion distribution coming
from the nucleon final state. Namely the  dominant contribution to it is the term proportional to $A_{qq}$ in
Eq.~(\ref{nsum}) with:
\ba
&& \left(2 (p-\Sigma_N^v u)\cdot q\,(p'-\Sigma_N^v u)\cdot q -
q^2\, [M_*^2+(p'-\Sigma_N^v u)\right. \nonumber \\
&& \left. \cdot (p-\Sigma_N^v u)]\right)=-2M_*^2 q^2,
\ea
which is the same expression as for the free nucleon, except that $M_*$ appears instead of $M$. 

Since $M_*/M<1$ a  further suppression  in addition to that from the Pauli blocking is obtained, depending on the actual value of $M_*/M$. 
The latter is difficult to constrain since  observables generally are only sensitive to the combination $M_*+\Sigma_N^v$. 
Since our aim is to make comparison with experiments on finite nuclei (rather than nuclear matter) with an average density smaller than 
the saturation density we assume small values 
for the energy shift in the range zero to $\Sigma_N^v=0.04$GeV, corresponding to effective mass values in the  range of of $0.85\,$GeV and
$0.89\,$GeV. These values are close to the ones used in more elaborate treatments of nuclear matter \cite{Jaminon:1989,Horowitz:1993}
where values of $0.8--0.85\,$GeV at saturation density give good agreement with observables.
The mean-field
shifts of the isobar mass and energy are chosen in such a way so that they reproduce the isobar-nucleon mass difference used in
Ref.~\cite{RLK-2009}. This means  $\Sigma_\Delta^s=-0.05\,$GeV and $-0.1\,$GeV and zero for the energy
shift.

\begin{figure}
\includegraphics[width=9cm]{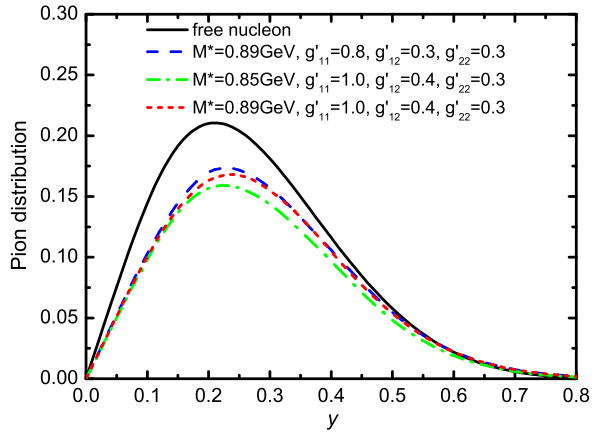}
\caption{ (Color online) Nucleon contribution to the pion distribution ($f^{\pi^0 N/A}(y)$) for the in-medium
nucleon compared to the case of free nucleon.}
\label{pinmed}
\end{figure}

\begin{figure}
\includegraphics[width=9cm]{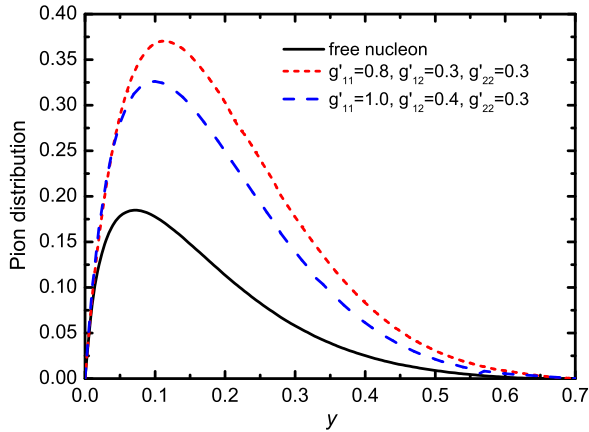}
\caption{ (Color online) Isobar contribution to the pion distribution ($f^{\pi^- \Delta/A}(y)$) in the in-medium
nucleon compared to the free nucleon. For both in-medium curves $M_*=0.89\,\text{GeV}$ and $\Sigma_N^v=0$.}
\label{pidmed}
\end{figure}

In Figs.~\ref{pinmed} and \ref{pidmed} we show the pion distributions $f^{\pi^0 N/A}(y)$ and $f^{\pi^- \Delta/A}(y)$
for in-medium nucleons for different parameter sets. For the nucleonic distribution one  observes a reduction coming partly
from the Pauli blocking of the nucleons in the medium and partly from the $M_*$ effect (which leads to a suppression
roughly by the factor $(M_*/M)^2$). The pion broadening in the nuclear medium only partly
compensates these effects and a net reduction is the result. This is not completely surprising since the
computations of Ref.~\cite{KLR-2009} do not lead to appreciable softening of the pion spectrum in the
medium which would result in enhanced pion distribution. In this respect the pion dressing of Ref.~\cite{KLR-2009}
is not significantly different from an older calculation \cite{KM-1995} which used a nonrelativistic  treatment of the isobar
and a softer pion-nucleon-delta form factor.
On the other hand a significant enhancement is observed for the contribution originating from the transition
$\text{N}\rightarrow \pi\Delta$ which is not Pauli suppressed. These results emphasize the importance of
careful treatment of the in-medium isobar self energy and propagator, which is made possible by the
convenient complete basis introduced in this context in Ref.~\cite{KD-2004}. 
As a consequence in the nuclear medium the combined effects from the pion and
from the isobar  can produce sizeable increase in the pion light-cone distribution. The latter is constrained
to smaller light-cone-momentum ratio $y$ values because of kinematical effect of the
isobar-nucleon mass difference but can still
have significant effects on the DY cross-section ratio.

\begin{figure}
\includegraphics[width=9cm]{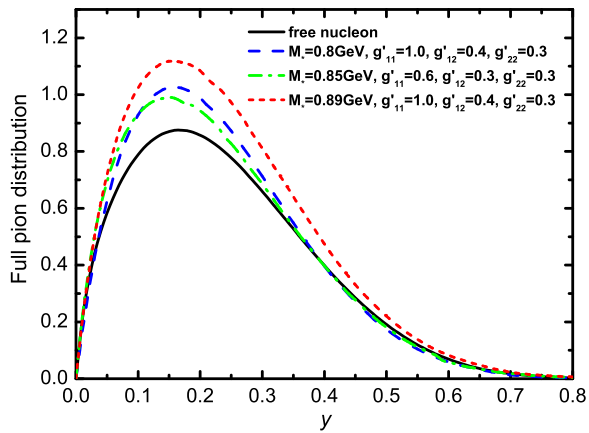}
\caption{ (Color online) Full pion distribution ($f^{\pi/A}(y)$) in the in-medium
nucleon compared to the case of free ``isoscalar" nucleon.}
\label{pifullmed}
\end{figure}

Since we are considering isospin symmetric nuclear medium and make a comparison with the deuteron it is
advantageous to consider the pion distribution in an ``isoscalar" nucleon, i.e.\ to consider a
proton-neutron average. Taking into account pions of all charges gives the complete pion distribution
of an ``isoscalar" nucleon:
\be
f^{\pi/A}(y)=3f^{\pi^0 N/A}(y)+2f^{\pi^- \Delta/A}(y).
\ee
In Fig.~\ref{pifullmed} we show the function $f^{\pi/A}(y)$ for diferent input parameter values compared
to the pion distribution of the free ``isoscalar" nucleon.
The probability $Z_A$ of the bare nucleon in the medium takes the values from 0.6 to 0.65, i.e. just slightly smaller than
in the free nucleon case.

\begin{figure}
\includegraphics[width=9cm]{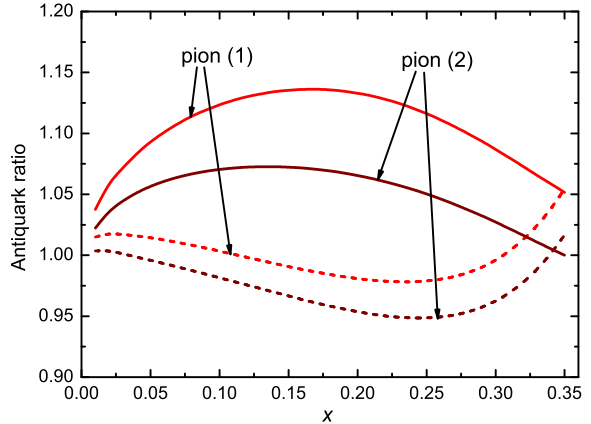}
\caption{ (Color online) The $\bar u^{\text{(medium)}}_p/\bar u^{\text{(free)}}_p$ (solid lines) and 
$\bar d^{\text{(medium)}}_p/\bar d^{\text{(free)}}_p$ (dash lines) ratios of antiquark distributions of an in-medium
proton relative to the free proton for parameter sets (1) and (2).}
\label{ubdbratio}
\end{figure}

Before examining the DY cross-section we show the ratio of the antiquark distribution in the in-medium proton and the 
same distribution in the free proton. The up and down antiquark distributions experience different in-medium 
modification due to different weights of nucleon and delta contributions even in isospin-symmetric nuclear medium.
In Fig.~\ref{ubdbratio} we show the ratios of antiquark distributions for an in-medium proton relative to the 
free one for two typical parameter sets denoted by ``pion (1)" and ``pion (2)", already used for plots 
in Figs.~\ref{termadd} and \ref{f2ratio}. 
We observe pronounced enhancement for the up antiquark coming from the substantial pion enhancement due to the 
delta-baryon final state as compared to quite modest enhancement and even suppression for the down antiquark 
as a consequence of larger weight of nucleon final state and smaller weight of delta-baryon final state as 
compared to the up antiquark. This difference points to the possibility of distinguishing between effects
coming from the medium modification of the nucleon and delta baryon by examining observables to which up and down
antiquarks contribute with different weights.

\begin{figure}
\includegraphics[width=9cm]{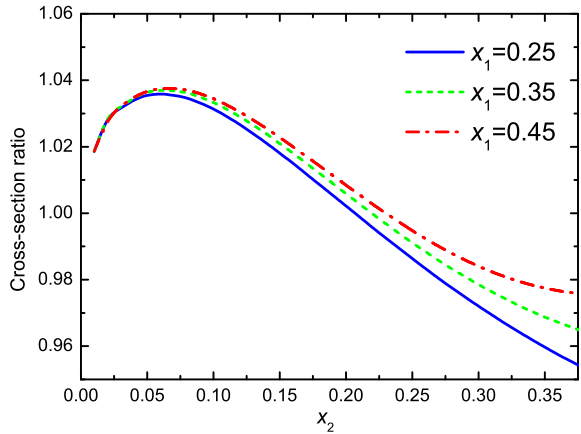}
\caption{ (Color online) The cross-section ratio (\ref{ratio}) as a function of $x_2$ for fixed
values of $x_1$. The used parameter values are: $M_*=0.85\,\text{GeV},\Sigma_N^v=0.04\,\text{GeV},
\Sigma_\Delta^s=-0.1\,\text{GeV},\Sigma_\Delta^v=0,g'_{11}=0.8,g'_{12}=0.3,g'_{22}=0.3$.}
\label{daratio-1}
\end{figure}

\begin{figure}
\includegraphics[width=9cm]{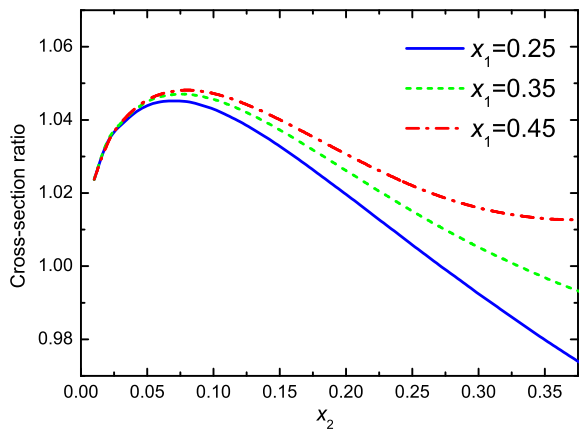}
\caption{ (Color online) The same as Fig.~\ref{daratio-1} but with
 used parameter values: $M_*=0.89\,\text{GeV},\Sigma_N^v=0,
\Sigma_\Delta^s=-0.1\,\text{GeV},\Sigma_\Delta^v=0,g'_{11}=1.,g'_{12}=0.4,g'_{22}=0.3$.}
\label{daratio-2}
\end{figure}

We now turn to the DY cross-section ratio (\ref{ratio}). In Figs.~\ref{daratio-1} and \ref{daratio-2} we show the
cross-section ratio (\ref{ratio}) as a function of $x_2$ for fixed
values of $x_1$. The input parameter values are given in the figure caption.
We observe an enhancement only for small values of $x_2$, typically less than 0.2, and for $x_2>0.1$
a decreasing trend as a result of the convolution with nucleon distribution (\ref{fermirel}).

\begin{figure}
\includegraphics[width=9cm]{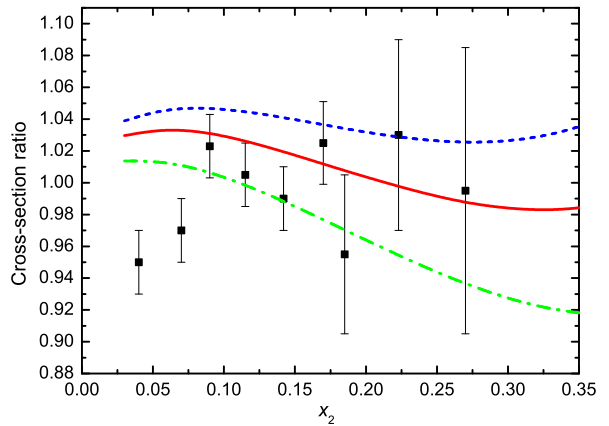}
\caption{ (Color online) Experimental results from Ref.~\cite{Alde90} compared to our calculation for
different parameter values. Short dash line: $M_*=0.89\,\text{GeV},\Sigma_N^v=0$; solid line: $M_*=0.85\,\text{GeV},
\Sigma_N^v=0.05\,\text{GeV}$; dash-dot line: $M_*=0.8\,\text{GeV},\Sigma_N^v=0.09\,\text{GeV}$. For all three curves:
$\Sigma_\Delta^s=-0.1\,\text{GeV},\Sigma_\Delta^v=0,g'_{11}=1.0,g'_{12}=0.4,g'_{22}=0.3$.}
\label{intratio-1}
\end{figure}

For comparison with the measurements of Ref.~\cite{Alde90} we computed the ratio of the nuclear
and deuteron cross sections for given $x_2$ and integrating over $x_1$ satisfying the condition
$x_1>x_2+0.2$ corresponding to the experimental cut-off. Fig.~\ref{intratio-1} shows the measured
values with error bars and the calculated curves for different input parameters.
We consider the lowest curve in Fig.~\ref{intratio-1} with $M_*=0.8\,$GeV and corresponding rather
pronounced suppression of the order $(M_*/M)^2$ probably exaggerating the effect of the nucleon
mean-field approximation and regard the other two curves as representing better our results based on the
preferred parameter sets.

\section{Summary}
In this work we presented an analysis of nuclear effects on the Drell-Yan process. The approach is based on the pion-cloud model of the nucleon
and a relativistically covariant self-consistent in-medium calculation of the pion and delta baryon propagators taking into account 
nuclear effects in the
mean-field approximation. Starting with the free nucleon we showed that the observed $\bar{d}-\bar{u}$ antiquark
distribution can be well reproduced by suitable choice of the $\pi\text{NN}$ and $\pi\text{N}\Delta$ form factors
with delta vacuum propagator taking into account its free width.

Using the same values for the form factors we
computed the pion light-cone-momentum distribution for nucleons in an isospin symmetric medium with density corresponding
to average densities of medium mass nuclei. We took into account the change of the pion cloud originating from both the
pion-nucleon and pion-delta states and a small correction (neglected in previous work) attributed to the binding effect of bare nucleon.

Fermi motion and binding of nuclear nucleons were accounted for by the two-parameter light-cone-momentum
distribution (\ref{fermirel}) which reproduces the negative slope of the classical EMC effect in the
region $0.1<x<0.5$ as shown in Fig.~\ref{f2ratio}. Taking into account the pion enhancement which comes from
the pion-delta state of the nucleon (which is significant only for small light-cone-momentum ($y\approx 0.2$)
values) leads to some enhancement of the $F_2(x)$ ratio for $x\leq 0.2$. 

Pion and delta properties in the nuclear medium are calculated in a recently developed fully covariant
self-consistent model \cite{KLR-2009} which consistently takes into account the $\pi\text{NN}$ and
$\pi\text{N}\Delta$ vertex corrections due to Migdal short-range correlations. Pronounced softening of the
in-medium pion spectrum present in simpler models does not appear in this approach and consequently Pauli
blocking causes some suppression of the pion distribution coming from the pion-nucleon state for an
in-medium nucleon. However, enhancement results from a careful treatment of the pion-delta state as a
consequence of pion broadening and delta shift and broadening.

 The net effect for preferred parameter
values is a modest enhancement of pion light-cone-momentum distribution mostly concentrated around
$y\approx 0.2$ value. As a consequence the DY cross-section ratio exceeds one for small values, typically
less than 0.2, of the $x_2$ variable for fixed $x_1$ values or integration over it corresponding to some
experimental cuts. The convolution with distribution (\ref{fermirel}) acts qualitatively on the antiquarks
in the same way as on the quarks producing a negative slope which is less pronounced for the integrated
cross section.

As one can see in Fig.~\ref{intratio-1} the large error bars of the measured cross-section ratio do
not allow a sensitive enough comparison with calculated results in order to determine more
precisely the preferred parameter values and consequently the in-medium properties of the pion and
delta baryon. It would be very desirable to achieve measurements with considerably smaller uncertainties
which could contribute to the resolution of some decades old issues of nuclear physics. 
We remark that another interesting possibility for studying sea quark distributions in nuclei would be the use of an  
Electron-Ion Collider as detailed in the joint report of the Brookhaven National Laboratory, the
Institute for Nuclear Theory (Seattle, WA) and the Thomas Jefferson National Accelerator Facility \cite{BNL/INT/JLAB}.
The proposed semi-inclusive deep-inelastic electron-nucleus scattering would provide new information about the structure
of nuclei and quantum chromodynamics of nuclear matter and extend possibilities for studying effects 
of the transverse momentum distribution of partons \cite{Gao:2010}.

\appendix*
\section{}
The terms of expression (\ref{dsum1}) necessary to calculate the contribution of the in-medium delta
are given as follows:
\begin{widetext}
\begin{eqnarray}
A_{qq}^{(\Delta)}&=&2\left|1+A_\Delta+q\cdot u\,B_\Delta/\sqrt{q^2-(q\cdot u)^2}\right|^2,\nonumber \\
A_{qu}^{(\Delta)}&=&-2\, {\text{Re}}\left[ q^2 B_\Delta/\sqrt{q^2-(q\cdot u)^2}(1+\bar A_\Delta -q\cdot u\,
\bar{B}_\Delta/\sqrt{q^2-(q\cdot u)^2})\right]
\nonumber \\
A_{uu}^{(\Delta)}&=& 2\left|q^2\,B_\Delta/\sqrt{q^2-(q\cdot u)^2}\right|^2,\nonumber \\
c_{qq}^{(Q)} &=& (M_*+(p-\Sigma_N^v u)\cdot \hat p')\left[ t+(q\cdot \hat p')^2-\left(q\cdot X(p')\right)^2\right],
\nonumber\\
c_{qq}^{(P)} &=& \frac{1}{3}(M_*+(p-\Sigma_N^v u)\cdot \hat p')\left[ t+(q\cdot \hat p')^2+3\left(q\cdot X(p')\right)^2\right],
\nonumber\\
c_{qu}^{(Q)} &=& -(M_*+(p-\Sigma_N^v u)\cdot \hat p')( q\cdot u\,-q\cdot \hat p'\,u\cdot \hat p'\,
+q\cdot X(p')\,u\cdot X(p')),\nonumber\\
c_{qu}^{(P)} &=& -\frac{1}{3}(M_*+(p-\Sigma_N^v u)\cdot \hat p')( q\cdot u\,-q\cdot \hat p'\,u\cdot \hat p'\,
-3q\cdot X(p')\,u\cdot X(p')), \nonumber\\
c_{uu}^{(Q)} &=& (M_*+(p-\Sigma_N^v u)\cdot \hat p')\left[ -1+(u\cdot \hat p')^2-(u\cdot X(p'))^2\right],
\nonumber\\
c_{uu}^{(P)} &=& \frac{1}{3}(M_*+(p-\Sigma_N^v u)\cdot \hat p')\left[ -1+(u\cdot \hat p')^2+3(u\cdot X(p'))^2\right],
\end{eqnarray}
with
\begin{eqnarray}
A_\Delta &\equiv &[\Pi^{(L)}g^{(L)}\,(1-\Pi^{(L)}g^{(L)})^{-1}]_{13}+
[\Pi^{(L)}g^{(L)}\,(1-\Pi^{(L)}g^{(L)})^{-1}]_{33}, \nonumber \\
B_\Delta &\equiv &[\Pi^{(L)}g^{(L)}\,(1-\Pi^{(L)}g^{(L)})^{-1}]_{23}+
[\Pi^{(L)}g^{(L)}\,(1-\Pi^{(L)}g^{(L)})^{-1}]_{43},\nonumber \\
X(p) &\equiv & \frac{(p\cdot u)\,p_\mu-p^2\,u_\mu}{p^2\,\sqrt{(p \cdot
u)^2/p^2-1}},
\end{eqnarray}
\end{widetext}
where $t=-q^2$ and $\hat p_\mu=p_\mu/\sqrt{p^2}$.

\begin{acknowledgments}
This research was supported in part by the Hungarian Research Foundation (OTKA)
grant 71989. CLK acknowledges the kind hospitality of the Kernfysisch Versneller
Instituut in Groningen.
\end{acknowledgments}

\bibliography{dy}

\end{document}